\documentclass[preprint2]{aastex}

\slugcomment{Submitted to Astrophysical Journal}

\shorttitle{O VII toward Markarian 421}
\shortauthors{Kaastra et al.}

\begin{document}


\title{The \ion{O}{7} X-ray forest toward Markarian~421:
Consistency between XMM-Newton and Chandra}


\author{J.S. Kaastra,  N. Werner and
J.W.A. den Herder}
\affil{SRON Netherlands Institute for Space Research,
Sorbonnelaan 2, 3584 CA Utrecht, The Netherlands}

\author{F.B.S. Paerels}
\affil{Columbia Astrophysics Laboratory,
Department of Astronomy, Columbia University,
550 West 120th Street, New York, 
NY 10027, USA}

\author{J. de Plaa\altaffilmark{1}}
\affil{SRON Netherlands Institute for Space Research,
Sorbonnelaan 2, 3584 CA Utrecht, The Netherlands}
 
\author{A.P. Rasmussen}
\affil{Kavli Institute for Particle Astrophysics and Cosmology, and Department
of Physics, Stanford University, CA 94305, USA}

\and

\author{C.P. de Vries}
\affil{SRON Netherlands Institute for Space Research,
Sorbonnelaan 2, 3584 CA Utrecht, The Netherlands}


\altaffiltext{1}{Astronomical Institute, Utrecht University, P.O. Box 80\,000, 3508 TA Utrecht,
The Netherlands}


\begin{abstract}
Recently the first detections of highly ionised gas associated with two Warm-Hot
Intergalactic Medium (WHIM) filaments have been reported. The evidence is based
on X-ray absorption lines due to \ion{O}{7} and other ions observed by Chandra
towards the bright blazar Mrk~421.
We investigate the robustness of this detection by a re-analysis of the original
Chandra LETGS spectra, the analysis of a large set of XMM-Newton RGS spectra of
Mrk~421, and additional Chandra observations. We address the reliability of
individual spectral features belonging to the absorption components, and assess
the significance of the detection of these components. We also use Monte Carlo
simulations of spectra. 
We confirm the apparent strength of several features in the Chandra
spectra, but demonstrate that they are statistically  not significant. This
decreased significance is due to the number of redshift trials that are made and
that are not taken into account in the original discovery paper. Therefore these
features must be attributed to statistical fluctuations. This is confirmed by
the RGS spectra, which have a higher signal to noise ratio than the Chandra
spectra, but do not show features at the same wavelengths. Finally, we show that
the possible association with a Ly$\alpha$ absorption system also lacks sufficient
statistical evidence. 
We conclude that there is insufficient observational proof for the existence of
the two proposed WHIM filaments towards Mrk~421, the brightest X-ray blazar on
the sky. Therefore, the highly ionised component of the WHIM still remains to
be discovered.
\end{abstract}



\section{Introduction}

Cosmological simulations show that the matter in the Universe is not uniformly
distributed but forms a web-like structure. Clusters of galaxies and
superclusters are found at the knots of this cosmic web. They are the location
of the highest mass concentration. The knots are connected through diffuse
filaments, containing a mixture of dark matter, galaxies and gas. According to
these simulations, the major part of this gas should be in a low density,
intermediate temperature ($10^5$ -- $10^7$~K) phase, the so-called Warm-Hot
Intergalactic Medium (WHIM, Cen \& Ostriker \cite{cen1999}). About half of all
baryons in the present day Universe should reside in this WHIM according to the
theoretical predictions, yet little of it has been seen so far. As the physics of
the WHIM is complicated and many processes are relevant to its properties,
observations of the WHIM are urgently needed. 

The density of the WHIM is low (typically 1 -- 1000 times the average baryon
density of the Universe (for example Dav\'e et al. \cite{dave2001}) or 0.3 --
300~m$^{-3}$) and the temperature is such that most emission will occur in the
EUV band. As the EUV band is strongly absorbed by the neutral hydrogen of our
Galaxy and its thermal emission is proportional to density squared, only the
hottest and most dense part of the WHIM can be observed in emission with X-ray
observatories such as XMM-Newton. Indications of \ion{O}{7} line emission from
the WHIM near clusters have been reported now (Kaastra et al.
\cite{kaastra2003}, Finoguenov et al. \cite{finoguenov2003}).

It is also possible to observe the WHIM in absorption, provided that a strong
background continuum source is present and provided a high-resolution
spectrograph is used. The first unambiguous detections of the WHIM in absorption
have been made in the UV band using bright quasar absorption lines from
\ion{O}{6} as observed with the Far Ultraviolet Spectroscopic Explorer (FUSE)
satellite (Tripp et al. \cite{tripp2000}; Oegerle et al. \cite{oegerle2000};
Savage et al. \cite{savage2002}; Jenkins et al. \cite{jenkins2003}).

Observation in the UV band is relatively easy because of the high spectral
resolution of FUSE and the relatively long wavelength $\lambda$ of the UV lines.
Since the optical depth at line centre scales proportional to the wavelength and
the best spectral resolution of the current X-ray observatories is an order of
magnitude poorer than FUSE, X-ray absorption lines are much harder to detect.

The first solid detection of X-ray absorption lines at $z=0$ towards a quasar
has been made by Nicastro et al. \cite{nicastro2002}. A spectrum of the bright
quasar \object{PKS~2155$-$304} taken with the Low Energy Transmission Grating
Spectrometer (LETGS) of Chandra showed zero redshift absorption lines due to
\ion{O}{7}, \ion{O}{8} and \ion{Ne}{9}. These lines have been confirmed by
XMM-Newton Reflection Grating Spectrometer (RGS) observations (Paerels et al.
\cite{paerels2003}; Cagnoni et al. \cite{cagnoni2003}), and similar X-ray
absorption lines have been seen towards several other sources, for example
\object{3C~273}, Fang et al. \cite{fang2003}; \object{NGC~5548}, Steenbrugge et
al. \cite{steenbrugge2003}; \object{NGC~4593}, McKernan et al.
\cite{mckernan2003}; and \object{Mrk~279}, Kaastra et al. \cite{kaastra2004}.
However, whether these lines are due to the local WHIM (for instance around our
Local Group) or that these lines have a Galactic origin is still debated.

Therefore, the first unambiguous proof of X-ray absorption lines from the WHIM
has to come from $z>0$ absorption lines. A first report of a $z=0.055$ X-ray
absorption system towards \object{PKS~2155$-$304} has been given by Fang et al.
\cite{fang2002} using a Chandra LETG/ACIS observation.  However, this feature
must be either a transient feature and therefore not associated to the WHIM, or
an instrumental artifact. This is because these results have not been confirmed
in both the more sensitive RGS observations towards this source (Paerels et al.
\cite{paerels2003}; Cagnoni et al. \cite{cagnoni2003}), nor in a long exposure
using the LETG/HRC-S configuration. The X-ray detection of 6 intervening UV
absorption systems towards \object{H~1821$+$643} in a 500~ks Chandra observation
by Mathur et al. \cite{mathur2003} should best be regarded as an upper limit due
to the rather low significance of these detections.

An apparently more robust detection of $z>0$ X-ray absorption due to the WHIM
has been presented by Nicastro et al. (\cite{nicastro2005a},
\cite{nicastro2005b}, NME herafter). These authors observed the brightest blazar
on the sky, \object{Mrk~421}, when the source was in outburst, typically an
order of magnitude brighter than the average state of this source. They obtained
two high-quality Chandra LETGS spectra of Mrk~421. They report the detection of
two intervening X-ray absorption systems towards this source in the combined
spectra. The first one at $z=0.011$ with a significance of 5.8$\sigma$ was
associated with a known \ion{H}{1} Ly$\alpha$ system, and the second one at
$z=0.027$ with a significance of 8.9$\sigma$ was associated with an intervening
filament $\sim 13$~Mpc from the blazar. 

The cosmological implications of the detection of the first X-ray forest are
perhaps the most important clue to resolving the problem of the "missing
baryons" (see Nicastro et al. \cite{nicastro2005a}), and therefore the
robustness of this detection is extremely important. Mrk~421 has been observed
many times by XMM-Newton as part of its routine calibration plan. The total
integration time of 950~ks accumulated over all individual observations with a
broad range of flux levels yields a spectrum with a quality superior to the
Chandra spectra taken during outburst. With such a high signal to noise ratio of
the spectra all kinds of systematic effects become important and we have
developed a novel way to analyse the RGS spectra taking account of these
effects. This new analysis is presented in the accompanying paper (Rasmussen et
al. \cite{rasmussen2006}). Here we present a carefull re-analysis of all
archival Chandra spectra of this source and compare the results from both
instruments.

\section{Observations and data analysis\label{sect:obs}}

\subsection{Analysis of the LETGS spectra\label{sect:letgsanalysis}}

We processed all 7 Chandra LETGS observations of Mrk~421 with a total exposure
time of 450~ks (see Table~\ref{tab:chandraobs}).  Two deep ($\sim100$~ks)
Chandra LETG/ACIS and LETG/HRC-S observations of Mrk~421 were triggered after
outbursts catching the source at its historical maximum. The $\sim20$~ks
LETG/HRC-S observation was obtained in Chandra Director Discretionary Time after
2 weeks of intense activity. The remaining four LETG/ACIS observations were
performed for calibration purposes. Our processed spectra contain in total
$\sim10^7$ counts. There are also two short observations with the High Energy
Transmission Grating Spectrometer (HETGS) of Chandra, but at the relevant
wavelengths ($\lambda > 20$~\AA) the signal to noise ratio of these spectra is
too low for our present investigation.

\begin{table*}[!ht]
\caption[]{Chandra LETGS observations of \object{Mrk~421}. The column "Set"
gives the number of the spectral data set into which each observation was
combined. The column "Flux" gives the flux at 20~\AA\ in
phot\,m$^{-2}$\,s$^{-1}$\,\AA$^{-1}$.}
\label{tab:chandraobs}
\centering
\begin{tabular}{clllcr}
\hline\hline
Set & Date & Obs.  & Detector  & Exposure & Flux   \\
    &      &  ID   &           &  (ks)    &  \\
\hline
1 & 2000 May 29 & 1715 & HRC-S   & 19.84 &  93 \\
2 & 2002 Oct 26 & 4148 & ACIS-S  & 96.84 & 333 \\
1 & 2003 Jul 01 & 4149 & HRC-S   & 99.98 & 242 \\
3 & 2004 May 06 & 5318 & ACIS-S  & 30.16 & 256 \\
3 & 2004 Jul 12 & 5331 & ACIS-S  & 69.50 &  39 \\
3 & 2004 Jul 13 & 5171 & ACIS-S  & 67.15 & 163 \\
3 & 2004 Jul 14 & 5332 & ACIS-S  & 67.06 & 140 \\ 
\hline
\end{tabular}
\end{table*}

The LETG/ACIS spectra were processed with CIAO~3.2.2 (CALDB~2.27) using the
standard Chandra X-ray Center pipeline. The spectra and ARFs were combined using
the CIAO tool $add\_grating \_spectra$. The LETG/HRC-S spectra were processed as
described in detail in Kaastra et al.\cite{kaastra2002}. 

We combined these seven spectra into three datasets as indicated in
Table~\ref{tab:chandraobs}: 
\begin{enumerate}
\item the $\sim100$~ks LETG/ACIS observation performed in 2000 when the source
had an exceptionally high luminosity (the spectrum contains $\sim4.3$ million
photons)
\item the combined two datasets obtained with LETG/HRC-S in 2000 and 2003 (the
spectrum contains $\sim2.7$ million photons)
\item the combined dataset of 4 calibration observations with LETG/ACIS taken in
2004, when the source was on average at a lower luminosity (the spectrum
contains $\sim3.2$ million counts). 
\end{enumerate}

We first analysed spectra 1 \& 2, which are the data sets analysed by NME.
However, contrary to NME we did not combine these data obtained in different
instrument configurations (LETG/ACIS and LETG/HRC-S) and rather decided to fit
the spectra simultaneously.  For the spectral fitting we used the SPEX package
(Kaastra et al. \cite{kaastra1996}). The two spectra obtained with different
instrument configurations were fitted with different continuum models but with a
common redshift and common absorption components. Since we are interested in
weak absorption features and not in the continuum emission of the source, we
fitted the underlying continuum with a spline model, using 12 grid points
between 9.5--35~\AA, with which we also removed any remaining broad instrumental
features (for more details on the \emph{spline} model see the SPEX
manual\footnote{see http://www.sron.nl/divisions/hea/spex/version2.0/release/
index.html}). This spectral range for fitting was chosen because it includes all
possible relevant spectral line features from neon, oxygen, nitrogen and carbon.

We modelled the Galactic absorption using the \emph{hot} model of the SPEX
package, which calculates the transmission of a plasma in collisional ionisation
equilibrium with cosmic abundances. The transmission of a neutral plasma was
mimicked by putting its temperature to 0.5 eV. We fix the Galactic absorption to
$1.61\times10^{24}$~m$^{-2}$ (Lockman \& Savage \cite{lockman1995}). We fit the
absorption lines of the $z=0$ local warm absorber, which is associated with the
ISM of our Galaxy or the Local Group WHIM filament, also with the \emph{hot}
model of the SPEX package, but now with the temperature as a free parameter. Our
best fit temperature of the local warm absorber is 0.07 keV with a hydrogen
column density of $7.4\times10^{22}$~m$^{-2}$. 

\begin{figure}[!ht]
\figurenum{1}
\includegraphics[angle=-90,width=\columnwidth]{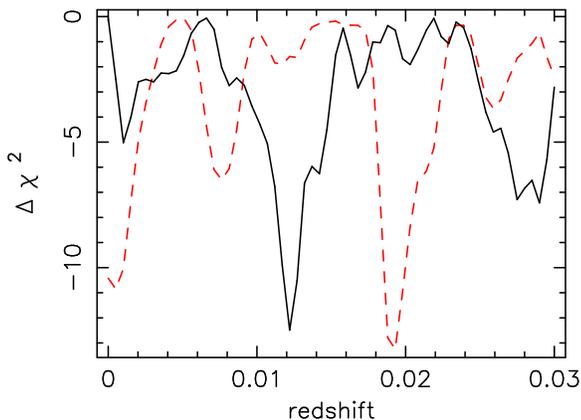}
   \caption{Improvement $\Delta\chi^2$ of the fit with weak absorption lines in the
model with respect to a fit without these absorption lines, as a function of
redshift.  Ions included are \ion{C}{6}, \ion{N}{6}, \ion{N}{7},
\ion{O}{6}, \ion{O}{7}, \ion{Ne}{9}, and \ion{Ne}{10}. The solid and dashed
lines represent the $\Delta\chi^2$ distribution as a function of  redshift when
fitting the observed spectrum and the simulated spectrum respectively.}
      \label{fig:deltachi}
\end{figure}

After obtaining a good fit of the continuum and the local warm absorber we
searched for the presence of weak absorption lines in the spectrum.  We searched
for redshifted absorption lines from \ion{C}{6}, \ion{N}{6}, \ion{N}{7},
\ion{O}{6}, \ion{O}{7}, \ion{O}{8}, \ion{Ne}{9}, \ion{Ne}{10} in all
possible WHIM filaments between our Galaxy and Mrk~421. We performed a
systematic search for these absorption features. Absorption lines of the above
mentioned ions were placed into one of 60 bins corresponding to different
redshifts between $z=0$ and $z=0.03$. In each bin the individual column
densities of the ions were fitted and the $\Delta\chi^{2}$ of the fit relative
to the fit without the weak absorption lines in the model was evaluated. To fit
the column densities of the ions we used the \emph{slab} model of the SPEX
package. That model calculates the transmission of a thin slab of matter with
arbitrary ionic composition. Free parameters are the velocity broadening
$\sigma_{\mathrm v}$ which we kept fixed to 100~km\,s$^{-1}$, and the ionic
column densities. Note that in our case the column densities are low so we
expect to be in the linear part of the curve of growth, such that
$\sigma_{\mathrm v}$ is irrelevant for the determination of the equivalent width
of the absorption lines. In order to really detect weak absorption lines we
performed this search with a spacing of 10~m\AA, one fifth of the resolution of
the LETGS (50 m\AA). Afterward we simulated a Chandra spectrum with the same
exposure time, same continuum model but without the weak absorption features and
again performed the same search on the simulated spectrum. The results of the
search on the real and on the simulated spectrum are shown in
Fig.~\ref{fig:deltachi}. 

We did not include redshifted \ion{O}{8} lines, since a part of the spectrum
where they may be present is influenced by small instrumental uncertainties. In
particular a nearby node boundary in LETG/ACIS with small-scale effective area
uncertainty would otherwise systematically bias the fit (see
Sect.~\ref{sect:individual} for more details).

\begin{figure}[!ht]
\figurenum{2}
\includegraphics[angle=-90,width=\columnwidth]{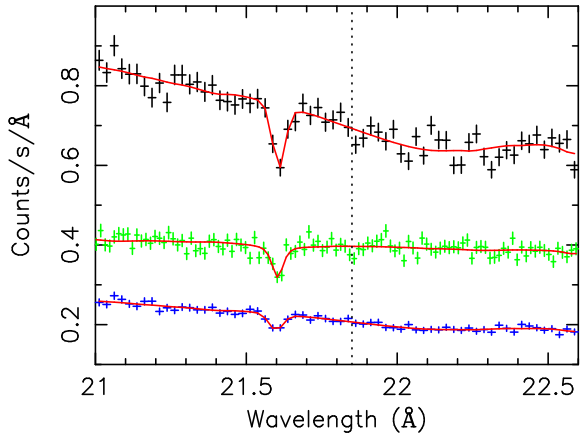}
\caption{
The $21.0 - 22.6$~\AA\ part of the spectrum of Mrk~421 with its best-fitting
continuum model plus an absorption model for the $z = 0$ absorber (solid line).
The spectrum at the top is the LETG/ACIS observation obtained when the source
was in an exceptionally bright state; the spectrum in the middle is the
LETG/HRC-S observation obtained in a bright state combined with a short
LETG/HRC-S observation obtained at a lower luminosity state; these observations
were combined and analysed by NME. The spectrum at the bottom was obtained
during 4 calibration observations. The two LETG/ACIS spectra were multiplied by
a factor of 1.5 for clarity of display. The dotted line shows the position of
the redshifted \ion{O}{7} line observed by NME.}
\label{compare}
\end{figure}

The most significant individual spectral feature which NME interpret as an
absorption line from a WHIM filament is the feature associated with an
\ion{O}{7} absorption line at z=0.011. To investigate that line further, we
considered all three datasets 1--3. We show the $21.0-22.6$~\AA\ spectral
interval extracted from these datasets in Fig.~\ref{compare}. The best-fitting
continuum model plus an absorption model for the local warm absorber is shown
with a solid line and the position of the redshifted O~VII line observed by NME
is indicated by the vertical dotted line. The spectral feature at 21.85~\AA\ is
seen in the spectrum obtained by LETG/HRC-S and a somewhat weaker and shifted
feature is seen in the LETG/ACIS spectrum obtained at the high source
luminosity. We note that NME combined these two datasets. In the third dataset
we do not see any feature at the indicated wavelength, despite the fact that the
spectrum contains more photons than the LETG/HRC-S observation. 

Finally, we determined the nominal redshift of the $z=0.011$ component from the
data sets 1 and 2, using only the strongest absorption lines (\ion{O}{7}
1s--2p). For the $z=0$ \ion{O}{7} line, we measure a line centroid of
$21.605\pm 0.006$ and $21.604\pm 0.005$~m\AA\ for spectrum 1 and 2,
respectively; for the $z=0.011$ components these values are  $21.864\pm 0.014$
and $21.854\pm 0.011$~m\AA. Therefore, the redshifts of the longer wavelength
line with respect to the $z=0$ line is $0.0121\pm 0.0007$ and $0.0117\pm
0.0005$, respectively. The weighted average is $z=0.0118\pm 0.0004$, somewhat
higher than the value obtained by NME.

\subsection{Analysis of the RGS spectra\label{sect:rgs}}

\begin{figure*}[!ht]
\figurenum{3}
\includegraphics[width=0.8\textwidth]{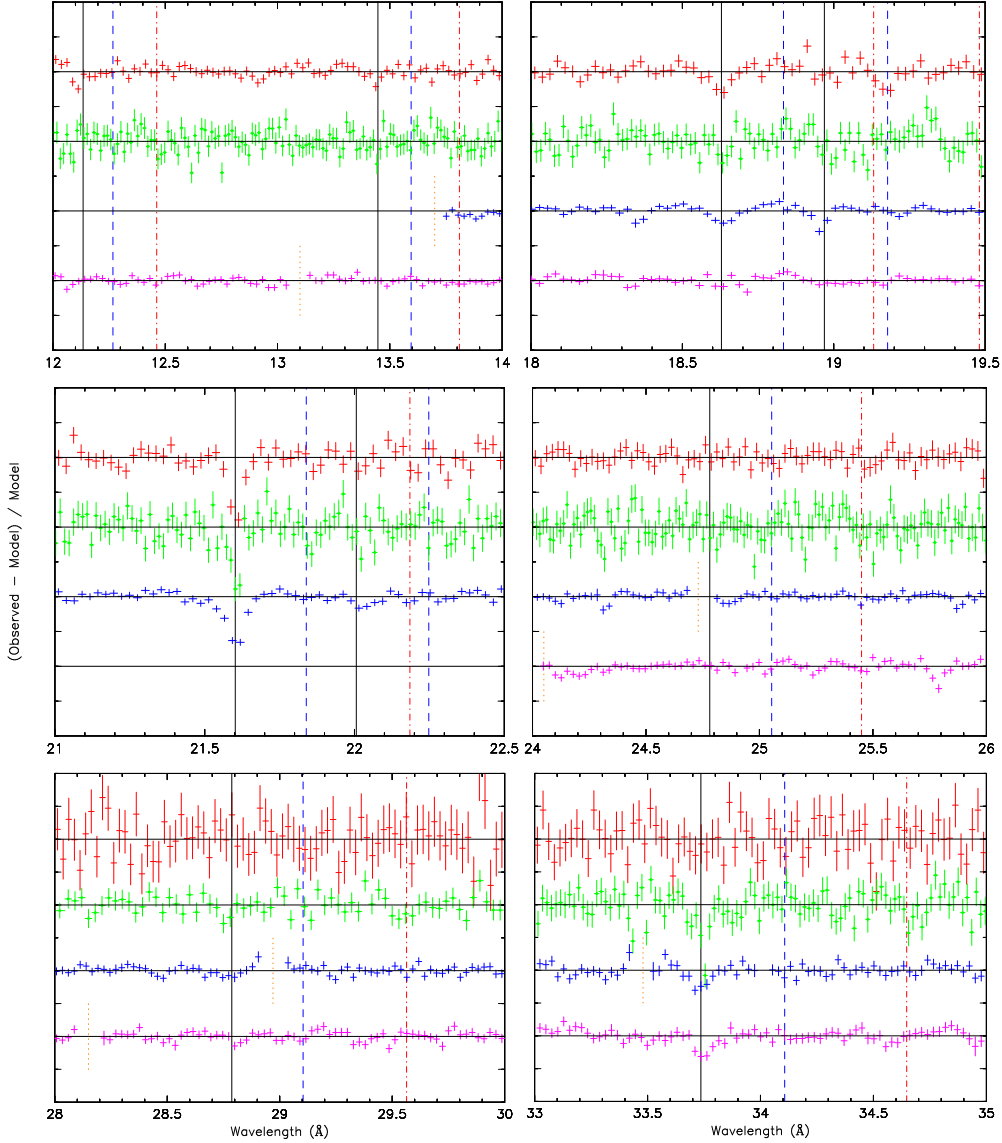}
   \caption{Comparison of the LETGS and RGS spectra of Mrk~421 in the 
   \ion{Ne}{10}/\ion{Ne}{9}, \ion{O}{8}, \ion{O}{7}/\ion{O}{6}, \ion{N}{7},
   \ion{N}{6} and \ion{C}{6} bands.  Each panel shows from top to bottom the
   LETG/ACIS spectrum (1), the LETG/HRC-S spectrum, RGS1 and RGS2. The
   quantities
   plotted are the fit residuals (observed$-$model)/model with respect to a
   local continuum model. One tickmark on the y-axis corresponds to 10~\%, and
   the data sets have been shifted arbitrarily along the y-axis in order to
   avoid overlapping data points. The solid vertical lines correspond to
   expected features at $z=0$, dashed lines to $z=0.011$, and dash-dotted lines
   to $z=0.027$. CCD gaps for RGS are shown with dotted lines in the appropriate
   spectrum. As explained by Rasmussen et al. \cite{rasmussen2006},  after
   careful screening occasionally an isolated instrumental dip may be present in
   a single RGS, such as at 25.8~\AA\ in RGS2.
	   }
      \label{fig:linecomp}
\end{figure*}

The analysis of the RGS spectra is described in detail by Rasmussen et al.
\cite{rasmussen2006}. We have taken their combined fluxed spectra and
corrected these for Galactic absorption using the same \emph{hot} model of SPEX
described above that was used for the LETGS spectra. We adjusted the oxygen
column density to $8.5\times 10^{-4}$ of the hydrogen column density in order to
get the best match around the oxygen edge. The absorption corrected spectra then
were fitted using a spline model with nodes separated by 0.5~\AA.  Obvious
absorption lines were ignored in this continuum fit. This procedure is needed
because the statistical quality of our combined data is so good: typically, the
flux in the oxygen region is determined with a statistical error of only 0.4~\%
per 0.5~\AA\ bin. There are remaining, large-scale uncertainties in the RGS
effective area of the order of a percent on \AA\ scales that would otherwise
bias the spectrum, and moreover to prove that the underlying continuum of all
these combined Mrk~421 observations would be a straight power law would be a
major challenge. Fig.~\ref{fig:linecomp} shows the residuals of this fit in the
same 6 wavelength intervals used by NME.

In addition, we show in Fig.~\ref{fig:linecomp} the fit residuals of the
LETGS spectra 1 and 2, after fitting with the same spline model with 
with nodes separated by 0.5~\AA\ as was used for the RGS data.

\subsection{Determing equivalent widths}

\begin{table*}[!ht]
\caption[]{The formal best-fit equivalent widths ($W_\lambda$) in m\AA\ for
absorption lines fixed at wavelengths of 21.60, 21.85, 22.20 and 22.02~\AA,
corresponding to the expected position of the  1s--2p lines of \ion{O}{7} at
$z=0$, $z=0.011$, $z=0.027$, and the 1s--2p doublet of \ion{O}{6} at $z=0$.
Negative values indicate emission lines. See the text for the details about the
LETGS datasets. We further list $W_\lambda$ as derived from the RGS spectra
(Sect.~\ref{sect:rgs}), and the weighted average for all three LETGS data sets
and the RGS data (row labelled w.a.). For comparison, we also list the equivalent
widths as determined by Williams et al. \cite{williams2005} and NME from the
combined LETGS sets 1 and 2 (labelled NME/W). All errors are r.m.s. errors
($\Delta\chi^2=2$). The last column gives an estimate
of the systematic uncertainty (also in m\AA) on the equivalent widths (not
included in the statistical errors).
}
\label{tab:ovii}
\centering
\begin{tabular}{@{}l@{$\quad$}r@{$\quad$}r@{$\quad$}r@{$\quad$}r@{$\quad$}r@{}}
\hline
\hline
Dataset & 21.60 & 21.85 & 22.20 & 22.02 & syst\\
\hline
LETGS 1 &$11.6\pm 1.3$& $1.5\pm 1.5$ & $3.1\pm 1.5$  & $1.9\pm 1.5$ & 0.6 \\
LETGS 2 &$11.7\pm 1.3$& $3.5\pm 1.4$ & $-1.5\pm 1.5$ & $1.1\pm 1.5$ & 0.9 \\
LETGS 3 &$10.3\pm 1.6$& $-0.6\pm 1.8$& $-0.4\pm 1.9$ & $0.1\pm 1.8$ & 0.6 \\
RGS     &$14.8\pm 0.7$& $0.4\pm 0.8$ & $-0.6\pm 0.7$  & $2.8\pm 0.7$ & 0.8 \\
\hline
w.a.    &$13.4\pm 0.5$& $1.0\pm 0.6$ & $-0.2\pm 0.6$  & $2.1\pm 0.6$ &\\
\hline
NME/W  & $9.4\pm 1.1$ & $3.0\pm 0.9$ & $2.2\pm 0.8$  & $2.4\pm 0.9$ &\\
\hline
\end{tabular}
\end{table*}

Equivalent widths of selected line features in the 21.0--22.5~\AA\ band were
determined from the fit residuals shown in Fig.~\ref{fig:linecomp}. We consider
the $z=0$ line of \ion{O}{7}, as well as the $z=0$ and $z=0.027$ \ion{O}{7}
lines and a 22.02~\AA\ feature discussed later. The equivalent widths
($W_\lambda$) are given in Table~\ref{tab:ovii}.

For RGS1 we used the exact line spread function. Following our earlier work
(Kaastra et al. \cite{kaastra2002}), we approximate the lsf of the LETGS by the
following analytical formula:
\begin{equation}
\phi \sim 1 / [1 + (\Delta\lambda/a)^2]^{\alpha},
\label{eqn:letgslsf}
\end{equation}
where $\alpha=2$ and $\Delta\lambda$ is the offset from the line centroid. The
scale parameter $a$ (expressed here in \AA) is weakly wavelength dependent as:
\begin{equation}
a = \sqrt{(0.0313)^2+(0.000306\lambda)^2}.
\end{equation} 
The errors on the equivalent widths that we list are simply the statistical 
errors on the best-fit line normalisation.
Table~\ref{tab:ovii} also shows the systematic uncertainties expected for each
line due to the uncertainty in the continuum level; these systematic
uncertainties were determined by shifting the continuum up and down by the 
1$\sigma$ statistical uncertainty in the average continuum flux for each local
0.5~\AA\  bin and re-determing the equivalent width.

Uncertainties in the shape of the lsf also may affect the derived values of
$W_\lambda$. We estimated this by modifying the lsf and determining how the
equivalent width of the 21.60~\AA\ line changes. For the LETGS, a change of
$\alpha$ from 2 (our preferred value) to 2.5 (the value used by the official
CIAO software) but keeping the same FWHM by adjusting $a$ leads to a decrease of
0.4~m\AA\ (3.5~\%)in $W_\lambda$; a decrease of $a$ by 10~\% leads to a decrease
of 0.5~m\AA\ (4.5~\%)in $W_\lambda$. If for the RGS instead of the true lsf a
Gaussian approximation would be used, the equivalent widths would be 20--30~\%
too small.  As shown by Rasmussen et al. \cite{rasmussen2006} the uncertainty
in the lsf gives slightly different results ($\sim 10$~\%) as two different
parameterisations of the lsf are used. For example, decreasing the FWHM of the
lsf of the RGS by 10~\% leads to a $W_\lambda$ decrease of 2.3~m\AA\ (15~\%).
Although the precise systematic uncertainty on $W_\lambda$ due to the lsf shape
is hard to establish accurately, it is most likely in the 5--15~\% range for
both LETGS and RGS. Taking this into account in addition to the statistical
uncertainty on $W_\lambda$ and the systematic uncertainty related to the
continuum level (Table~\ref{tab:ovii}), we conclude that there is not a large
discrepancy between the equivalent widths of the $z=0$ \ion{O}{7} line
measured by both instruments. 

\section{Discussion}

\subsection{Significance of the absorption components detected by Chandra}

The evidence for the two absorption components at $z=0.011$ (5.8$\sigma$) and
$z=0.027$ (8.9$\sigma$) as presented by  NME looks at a first glance convincing.
However, there are good reasons to put question marks to these significances. We
follow here two routes: first we discuss the significance by looking into detail
to the individual features identified by NME, and next we analyse the
significance of the absorption components in a coherent way.

\subsubsection{Individual line features\label{sect:individual}}

NME found 28 or 24 absorption features with a significance of more than
3$\sigma$ in their analysis of the combined LETG/HRC-S and LETG/ACIS spectrum.
The precise number is somewhat unclear from their paper but is not important
here. Of these 24 lines, 14 have an origin at $z=0$ and there is no doubt about
the significance of that component, so we focus on the 10 features with a
tentative identification with $z>0$ absorption lines. In total 3 of these belong
to the $z=0.011$ system, and 6 to the $z=0.027$ system. One feature at
24.97~\AA\ was unidentified by NME. We will discuss these features here
individually as we point out that not all of these detections can be regarded as
unambiguous.

The absorption component at $z=0.011$ had three features:
\begin{enumerate}
\item \ion{O}{8} 1s--2p at 19.18~\AA: Fig.~\ref{fig:linecomp} shows that
there is only evidence for an absorption feature from the LETG/ACIS spectrum,
not from the LETG/HRC-S or RGS spectra. The feature coincides exactly with a
$\sim$ 0.05~\AA\ wide dip of 3~\% depth in the effective area of the LETG/ACIS,
associated with an ACIS node boundary (as was also pointed out by NME); 
therefore this feature is most likely of instrumental origin.
\item \ion{O}{7} 1s--2p at 21.85~\AA: there is no clear feature in the RGS
spectrum, but there is a feature in both LETGS spectra. However, the centroid in
the LETG/ACIS spectrum seems to occur at slightly longer wavelength, by
$\sim$~0.01~\AA. There are no known instrumental features in the LETGS at this
wavelength, so this may be real provided it has sufficient significance. But see
Sect.~\ref{sect:selfcons}.
\item \ion{N}{7} 1s--2p at 25.04~\AA: no feature present in the RGS deeper
than 1~\%; the LETG/HRC-S spectrum (Fig.~\ref{fig:linecomp}) shows a $\sim 5$~\%
negative deviation at a slightly smaller wavelength, and the (here noisy)
LETG/ACIS spectrum may be consistent with this. The smaller wavelength is
consistent with the findings of NME, who give here a redshift of 0.010 instead
of 0.011.
\end{enumerate}
In summary, the X-ray evidence for the 0.011 component rests solely upon the
detection of two lines with different redshift and only seen in the LETGS.
As we will argue later, even the statistical significance of the combined
features is insufficient to designate it as a robust detection.

The component at $z=0.027$ had six features:
\begin{enumerate}
\item \ion{Ne}{9} 1s--2p at 13.80~\AA: this feature coincides with a narrow dip
in the effective area  of the LETG/ACIS associated with a node boundary, similar
to the $z=0.011$ \ion{O}{8} 1s--2p line. It is therefore no surprise that it
is only seen in the LETG/ACIS spectrum, with perhaps a little but not very
significant strengthening from the LETG/HRC-S spectrum. It is not visible in the
RGS spectrum. More importantly, its best-fit redshift as reported by NME of
0.026 is off from the redshift of this component at $z=0.027$. We conclude that
this feature has an instrumental origin.
\item \ion{O}{7} 1s--3p at 19.11~\AA: there is only a weak signal visible in
the LETG/HRC-S spectrum (Fig.~\ref{fig:linecomp}); this looks a little less
pronounced than in the plot given by NME but this may be due to different
binning; more importantly, given the weakness of the 1s--2p transition in
the same ion (see below) and the expected 5 times smaller equivalent width of
the 1s--3p line as compared to the 1s--2p line makes any detection of the 1s--3p
line impossible. 
\item \ion{O}{7} 1s--2p at 22.20~\AA: only clearly seen in the LETG/ACIS
spectrum. No evident instrumental features are present. However, the redshift of
0.028 is slightly off from the redshift of this component at $z=0.027$.
\item \ion{N}{7} 1s--2p at 25.44~\AA: a weak feature is present mainly in the
LETG/HRC-S spectrum; RGS1 may be consistent with this. Given the weakness of the
oxygen lines, a clear detection of \ion{N}{7} would be surprising unless the
absorber has an anomalously high nitrogen abundance.
\item \ion{N}{6} 1s--2p at 29.54~\AA: here the LETG/HRC-S spectrum has the best
statistics and it shows indeed a shallow dip with four adjacent bins
approximately 1$\sigma$ below the reference level. The same remark as above
about the nitrogen abundance can be made. Moreover, there is nothing there in
the RGS spectra.
\item \ion{C}{6} 1s--2p at 34.69~\AA: as above the feature is mainly evident
from the LETG/HRC-S line. As the cosmic carbon abundance is significantly higher
than the nitrogen abundance, this line is more likely to be detectable. Again,
the redshift of 0.028 is on the high side.
\end{enumerate}
In summary, the detection of the $z=0.027$ component rests upon \ion{O}{7},
seen only in the LETG/ACIS spectrum, and the 1s--2p lines of \ion{N}{7},
\ion{N}{6} and \ion{C}{6} as well as the 1s--3p lines of \ion{O}{7}, seen
only in the LETG/HRC-S spectrum. None of these features is visible in both LETGS
configurations, not all of these lines have exactly the same redshift and some
of the lines (\ion{O}{7} 1s--3p and the nitrogen lines) are problematic from a
physical point of view.

\subsubsection{Self-consistent assessment of the significance of the
line features\label{sect:selfcons}}

In the previous section we have put several question marks to the significance
and reliability of the nine $z=0.011$ and $z=0.027$ line identifications as
derived from the LETGS spectra by NME. Nevertheless, some of the features
identified by these authors indeed show negative residuals at the predicted
wavelengths for these components, with a nominal statistical significance as
determined correctly by NME.

However, the actual significance of these detections is much smaller, and
actually as we show below the redshift components are not significant at all. 
This is due to two reasons. 

First, NME fitted line centroids for each component separately, and then
determined the significance for each line individually. However, for all lines
belonging to the same redshift component, the wavelengths are not free
parameters but are linked through the value of the redshift. Although NME later
assess the significance of the components by fixing the wavelengths of the
relevant lines, they first have established the presence of redshifted
components and their redshift values by ignoring this coupling. Their argument
for "noise" in the wavelength solution of in particular the LETG/HRC-S detector
is not relevant here; first, our improved treatment of the LETGS gives smaller
noise, but more importantly, if there is noise, then for each potential line the
amplitude and direction of the noise are unknown, and therefore there is no
justification for adjusting the line centroids by eye or by chi-square: the true
wavelength alinearity might just have the opposite sign. The only way to do this
unbiased is to couple the lines from different ions even before having found the
redshifted components.

The most important problem is however the number of trial redshifts that have
been considered. NME started by looking to $>3\sigma$ excesses in the data in
order to identify the two redshift systems. However, a $>3\sigma$ excess has
only meaning for a line with a well-known wavelength based on \emph{a priori}
knowledge. For example, knowing that there is a highly ionised system with a
given redshift in the line of sight makes it justified to look for \ion{O}{8}
or \ion{O}{7} absorption lines at precisely that redshift. Here this a priori
knowledge is lacking (only \emph{a posteriori} circumstantial evidence based on
galaxy association or association \ion{H}{1} Ly$\alpha$ absorption is
presented). In a long enough stretch of data from any featureless spectrum there
will always be several statistical fluctuations that will surpass the $3\sigma$
significance level. The significance of any feature with unknown wavelength is
therefore not determined by the cumulative distribution $F(x)$ of the excess $x$
(with $F(x)$ usually a Gaussian) at a given wavelength. Instead it is given by
the distribution $G(x)$ of the maximum of $F$ for $N$ independent trials, which
is $G(x) = F^N(x)$. For a large number of trials, $F$ has to be very high to get
a significant value of $G$. 

The number of independent trials $N$ is of order of magnitude $W/\Delta\lambda$,
with $W$ the length of the wavelength stretch that is being searched, and
$\Delta\lambda$ the instrumental Full-Width at Half Maximum (FWHM). However,
this is only an order of magnitude estimate. The precise value depends on the
shape of the instrumental line spread function and the bin size of the data.

By using a Monte Carlo simulation we have assessed the significance of line
detections for the present context. We start from a simple featureless model
photon spectrum. This spectrum is binned with a bin size $\delta$ of 0.015~\AA.
For each spectral bin $i$, the deviations $d_i$ of the observed spectrum from
this model spectrum are normalised to their nominal standard deviations
$\sigma_i$ such that $x_i=d_i/\sigma_i$ has a standard normal distribution with
mean $0$ and variance $1$. Given the high count rate of the spectrum, the
approximation of the Poissonian distribution for each bin with a Gaussian as we
do here is justified. All $x_i$ are statistically independent random variates.
We consider 7 spectral lines, centred on $\lambda_j$ equal to 12.1, 13.4, 21.6,
22.0, 24.8, 28.8 and 33.7~\AA, and generate the $x_i$ for the wavelength range
between $\lambda_j$ and $\lambda_j(1+z_{\max})$ with $z_{\max}$ the redshift of
Mrk~421 which we approximate here by 0.03. Then we do a grid search over
redshift (step size 0.0003 in $z$) and for each redshift we determine the best
fit absorption line fluxes. The fluxes are simply determined from a least
squares fit taking account of the exact line profile $\phi(\lambda)$
(Eqn.~\ref{eqn:letgslsf}).

\begin{figure}
\figurenum{4}
\includegraphics[angle=-90,width=\columnwidth]{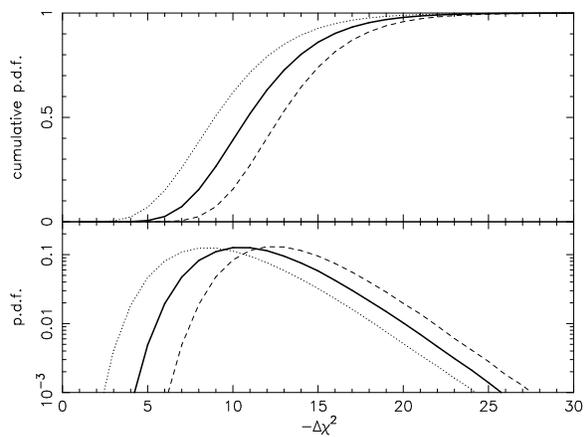}
   \caption{Probability density function (lower panel) and cumulative
   probability density function (upper panel) for the maximum improvement
   $-\Delta\chi^2$ in LETGS fits including absorption lines at
    12.1, 13.4, 21.6, 22.0, 24.8, 28.8 and 33.7~\AA, when a redshift range
    between $z=0$ and $z=z{_{\max}}$ is searched. Here results for 3 values
    of $z_{\max}$ are shown: dotted line: $z{_{\max}}=0.015$; thick solid
    line: $z{_{\max}}=0.03$ (appropriate for Mrk~421); dashed line:
    $z{_{\max}}=0.06$.
	   }
      \label{fig:sigplot}
\end{figure}

As we are looking for absorption lines, we put the normalisation of the line to
zero whenever it becomes positive. We then calculate the difference $\Delta$ in
$\chi^2$ between the model including any absorption lines found, to the model
without absorption lines. As the fit always improves by adding lines,
$\Delta\chi^2$ is always negative. For each run, we determine the
redshift with the most negative value of $\Delta\chi^2$, i.e. the most
significant signal for any redshift. We repeat this process $10^6$ times
and determine the statistical distribution of $\Delta\chi^2$ by this way
(see Fig.~\ref{fig:sigplot}).

Our results are almost independent of the chosen bin size $\delta$ for the
spectrum (we verified that by numerical experiments). In the context of
\object{Mrk~421}, the median and peak of the distribution are close to 10. This
corresponds to a "$5\sigma$" detection if no account is taken of the number of
trials. Components with apparent significances stronger than 5.8 or 8.9$\sigma$
as obtained by NME for the $z=0.011$ and $z=0.027$ systems, respectively, have
probabilities of 40 and 6~\% to occur by chance. Thus, the $z=0.011$ component
is not significant at all. Furthermore, as we show below in the next section,
the $z=0.027$ is not as significant as reported by NME.

\subsubsection{The significance of the line spectra from spectral simulations}

The two peaks at $z=0.011$ and $z=0.027$ are indeed reproduced by our new
analysis of the LETGS spectra (Fig.~\ref{fig:deltachi};
Sect.~\ref{sect:letgsanalysis}). While the peak at $z=0.011$ has a significance
similar to the value quoted by NME, the $z=0.027$ component is less significant.
This is mainly because we omitted the \ion{O}{8} wavelength range because of
the reasons mentioned earlier. Note that both troughs are not very narrow; this
corresponds to the slightly different redshifts found for the lines of NME.

We also took the same continuum spectrum and used it to make a simulation
(Fig.~\ref{fig:deltachi}). The simulated spectrum shows three peaks of similar
strength, at $z=0.001$, $z=0.007$, and $z=0.019$. In both the fitted true
spectrum of \object{Mrk~421} and the simulated spectrum, the strongest peak in
terms of $\Delta\chi^2$ is in excellent agreement with the predictions as shown
in  Fig.~\ref{fig:sigplot}. 

We conclude that based on the Chandra data alone there is not sufficient
evidence for a significant detection of the WHIM towards Mrk~421. This does of
course also imply that a true WHIM signal must be much stronger or more
significant than the detections of NME before it can be regarded as evidence for
the existence of the WHIM.

\subsection{Association of the $z=0.011$ component with a Ly$\alpha$ absorber.}

A strong argument in favour of the existence of the $z=0.011$ WHIM component is
its association with a Ly$\alpha$ absorption system. Penton et al.
\cite{penton2000} found an absorption line with an equivalent width of $86\pm
15$~m\AA\ at 1227.98~\AA, identified as \ion{H}{1} Ly$\alpha$ at a redshift of
$z=0.01012\pm 0.00002$. Savage et al. \cite{savage2005} have studied
extensively UV spectra of Mrk~421 obtained by FUSE and HST but they could not
find any other \ion{H}{1} lines neither any associated metal lines.  However,
Savage et al. argue that this line could be hardly anything else than
Ly$\alpha$. At the given redshift, the system should be a \ion{H}{1} cloud in a
galactic void.

From our analysis of the apparent $z=0.011$ system
(Sect.~\ref{sect:letgsanalysis}), we found as a best fit redshift $z=0.0118\pm
0.0004$. This is higher than the value quoted by NME because of our improved
wavelength solution of the LETGS. The difference of $500\pm 120$~km\,s$^{-1}$
between this X-ray redshift and the Ly$\alpha$ redshift corresponds to
0.036~\AA\ at the wavelength of the \ion{O}{7} resonance line. This is almost
equal to the spectral resolution of the LETGS at that wavelength and definitely
inconsistent with our current understanding of the LETGS wavelength scale. 

We conclude that there is no evidence for a direct physical connection between
the putative $z=0.011$ X-ray absorption component and the Ly$\alpha$ absorber. A
similar conclusion was also reached by Savage et al. \cite{savage2005}, based
on the narrowness of the Ly$\alpha$ line which implies a low temperature and the
absence of any \ion{O}{6} absorption in the FUSE spectra. Even the evidence for
a more loose association of both systems, for example in different parts of a
larger filamentary structure, is not convincing: over the redshift range of
$0\le z \le 0.03$, the probability that the most significant statistical X-ray
fluctuation would occur within $\vert \Delta z \vert <0.0017$ from the only
known intervening Ly$\alpha$ system is 11.2~\%.

\subsection{The final argument: no lines in the RGS spectrum}

We have shown above that based on the Chandra data alone there is insufficient
proof for the existence of the $z=0.011$  and $z=0.027$ absorption components
reported by NME. The lack of evidence does not imply automatically that
these components are not present, but just that their presence cannot be
demonstrated using LETGS data.

However, if these components really would exist at the levels as reported by
NME, then adding more data like our combined RGS data set should enhance the
significance, but the opposite is true: the RGS spectra show no evidence at all
for both absorption components. Taking the weighted average of the equivalent
widths of the \ion{O}{7} lines obtained by all instruments, we find for the
$z=0.011$ and $z=0.027$ components equivalent widths of $1.0\pm 0.6$ and
$-0.2\pm 0.6$~m\AA, respectively (see Table~\ref{tab:ovii}), to be compared to
$3.0\pm 0.9$ and $2.2\pm 0.8$ as reported by NME. Hence, the column densities of
any component are at least two times smaller than reported by NME.

The lack of evidence for both components in RGS spectra was also reported by
Ravasio et al. \cite{ravasio2005}, based on a smaller subset of only 177~ks.
Our analysis uses 4 times more exposure time. 

Finally, while we were writing our paper, a preprint by Williams et al.
\cite{williams2006} appeared. These authors use 437~ks RGS data, only half of
our exposure time. They also do not find evidence for the two absorption
systems, but attribute this to "narrow instrumental features, inferior spectral
resolution and fixed pattern noise in the RGS". As we have shown in the
accompanying paper (Rasmussen et al. \cite{rasmussen2006}) these statements lack
justification when a careful analysis is done.

\subsection{The 22.02~\AA\ feature}

Both the RGS and the LETGS spectra seem to indicate the presence of  an
absorption line at 22.02~\AA, with combined equivalent width of $2.1\pm
0.6$~m\AA\ (Table~\ref{tab:ovii}). This feature was identified as $z=0$
absorption from \ion{O}{6}, based upon the almost exact coincidence with the
predicted wavelength of that line (Williams et al. \cite{williams2005}).
However, as  Williams et al. point out, the equivalent width is about 3 times
higher than the value predicted from the 2s--2p doublet at 1032 and 1038~\AA.
Williams et al. then offer three possible explanations for the discrepancy.

The first explanation, wrong oscillator strengths of the K-shell transitions by
a factor of 2--4, can be simply ruled out. These strong transitions can be
calculated with much higher precision. 

The second explanation invokes a large fraction of \ion{O}{6} in an excited
state. This would suppress the 2s--2p UV lines relative to the 1s--2p X-ray
lines. Apart from the fact that this situation is hard to achieve in a low
density plasma (one needs almost LTE population ratio's), there is another
important argument against this. The X-ray absorption lines from the 1s$^2$\,2p
configuration are shifted by 0.03--0.05~\AA\ towards longer wavelength as
compared to lines from the ground state 1s$^2$\,2s (A.J.J. Raassen, private
communication). Thus, in this scenario the X-ray line should shift by a
measurable amount towards longer wavelengths, which is not observed.

The third explanation offered by Williams et al. invokes intervening
\ion{O}{7} at $z=0.0195$. Given the lack of other absorption lines from the
same system, Williams et al. argue that this is unlikely.

Our strongest argument against a real line is however its significance. Using
the known column density derived from the FUSE spectra, $(2.88\pm 0.14) \times
10^{18}$~m$^{-2}$, Savage et al. \cite{savage2005}), we estimate that the
corresponding $z=0$ 1s--2p line of \ion{O}{6} should have an equivalent width
of 0.64~m\AA. Subtracting this from the observed equivalent width, the
"unexplained" part has an equivalent width of $1.5\pm 0.6$~m\AA, i.e. a
2.5$\sigma$ result. However, similar to our analysis of the  Chandra data, it is
easy to argue that given the number of redshift trials this significance is
insufficient to be evidence for redshifted \ion{O}{7}.

\section{Conclusions}

We have re-analysed carefully the Chandra LETGS spectra of Mrk~421, the first
source for which the detection of the X-ray absorption forest has been claimed.
We have supplemented this with an analysis of additional LETGS observations as
well as with a large sample of XMM-Newton RGS observations, amounting to a total
of $950$~ks observation time. We find that there is no evidence for the presence
of the $z=0.011$ and $z=0.027$ filaments reported by NME. Also the association
with an intervening \ion{H}{1} Ly$\alpha$ absorption system is not sufficiently
supported.


\begin{thebibliography}{}

\bibitem[2003]{cagnoni2003}
Cagnoni, I., Nicastro, F., Maraschi, L., Treves, A., \& Tavecchio, F., 2003, 
New Astr. Rev., 47, 561

\bibitem[1999]{cen1999} 
Cen, R., \& Ostriker, J.P., 1999, ApJ, 514, 1

\bibitem[2001]{dave2001}
Dav\'e, R., Cen, R., Ostriker, J.P., et al., 2001, ApJ, 552, 473

\bibitem[2001]{denherder2001}
den Herder, J.W., Brinkman, A.C., Kahn, S.M., et al., 2001, A\&A, 365, 1

\bibitem[2002]{fang2002}    
Fang, T., Marshall, H.L., Lee, J.C., Davis, D.S., \& Canizares, C.R.,  
2002, ApJ, 572, L127

\bibitem[2003]{fang2003}
Fang, T., Sembach, K.R., \& Canizares, C.R., 2003, ApJ, 586, L49

\bibitem[2003]{finoguenov2003}
Finoguenov, A., Briel, U.G., \& Henry, J.P., 2003, A\&A, 410, 777

\bibitem[2003]{jenkins2003}
Jenkins, E.B., Bowen, D.V., Tripp, T.M., et al., 2003, AJ, 125, 2824

\bibitem[1996]{kaastra1996}
Kaastra, J.S., Mewe, R., \& Nieuwenhuijzen, H., 1996, 
in Frontiers Science Ser. 15, 
UV and X-ray Spectroscopy of Astrophysical and Laboratory Plasmas,
Eds.\ K. Yamashita \& T. Watanabe (Univ. Ac. Press, Tokyo) 411

\bibitem[2002]{kaastra2002}
Kaastra, J.S., Steenbrugge, K.C., Raassen, A.J.J., et al., 2002, A\&A, 386, 427

\bibitem[2003]{kaastra2003}
Kaastra, J.S., Lieu, R., Tamura, T., Paerels, F.B.S., \& den Herder, J.W.,
2003, A\&A, 397, 445

\bibitem[2004]{kaastra2004}
Kaastra, J.S., Raassen, A.J.J., Mewe, R., et al., 2004, A\&A, 428, 57

\bibitem[1995]{lockman1995}
Lockman, F.J. \& Savage, B.D., 1995, ApJS 97, 1

\bibitem[2003]{mathur2003}
Mathur, S., Weinberg, D.H., \& Chen, X., 2003, ApJ, 582, 82

\bibitem[2003]{mckernan2003}
McKernan, B., Yaqoob, T., George, I.M., \& Turner, T.J., 2003, ApJ, 593, 142

\bibitem[2002]{nicastro2002}
Nicastro, F., Zezas, A., Drake, J., et al., 2002, ApJ, 573, 157

\bibitem[2005a]{nicastro2005a}
Nicastro, F. Mathur, S., Elvis, M., et al., 2005a, Nature, 433, 495

\bibitem[2005b]{nicastro2005b}
Nicastro, F. Mathur, S., Elvis, M., et al. (NME), 2005a, ApJ, 629, 700

\bibitem[2000]{oegerle2000}
Oegerle, W.R., Tripp, T.M., Sembach, K.R., et al., 2000, ApJ, 538, L23

\bibitem[2003]{paerels2003}
Paerels, F.B.S., Rasmussen, A., Kahn, S., den Herder, J.W., \& de Vries, C.P.,  
2003, MPE Report, 281, 57

\bibitem[2000]{penton2000}
Penton, S.V., Shull, J.M., \& Stocke, J.T., 2000, ApJ, 544, 150

\bibitem[2006]{rasmussen2006}
Rasmussen, A.P., de Vries, C.P., Paerels, F.B.S., et al., 2006, ApJ, submitted.

\bibitem[2005]{ravasio2005}
Ravasio, M., Tagliaferri, G., Pollock, A.M.T., Ghisellini, G., \& Tavecchio, F.,
2005, A\&A, 438, 481

\bibitem[2002]{savage2002}
Savage, B.D., Sembach, K.R., Tripp, T.M., \& Richter, P., 2002, ApJ, 564, 631

\bibitem[2005]{savage2005}
Savage, B.D., Wakker, B.P., Fox, A.J., \& Sembach, K.R., 2005, ApJ, 619, 863

\bibitem[2003]{steenbrugge2003}
Steenbrugge, K.C., Kaastra, J.S., de Vries, C.P., \& Edelson, R., 2003, 
A\&A, 402, 477

\bibitem[2000]{tripp2000}
Tripp, T.M., Savage, B.D., \& Jenkins, E.B., 2000, ApJ, 534, L1

\bibitem[2005]{williams2005}
Williams, R.J., Mathur, S., Nicastro, F., et al., 2005, ApJ, 631, 856

\bibitem[2006]{williams2006}
Williams, R.J., Mathur, S., Nicastro, F., \& Elvis, M., 2006, submitted to ApJ
(astro-ph/0601620)

\end{thebibliography}
\end{document}